\documentclass[preprint,showpacs,preprintnumbers,amsmath,amssymb,natbib]{revtex4}

\usepackage{graphicx}
\usepackage{epsfig}		
\usepackage{dcolumn}
\usepackage{bm}
\def\0{\mbox{\tiny $0$}}
\def\1{\mbox{\tiny $1$}}
\def\2{\mbox{\tiny $2$}}
\def\3{\mbox{\tiny $3$}}
\def\4{\mbox{\tiny $4$}}
\def\5{\mbox{\tiny $5$}}
\def\6{\mbox{\tiny $6$}}
\def\7{\mbox{\tiny $7$}}
\def\8{\mbox{\tiny $8$}}
\def\9{\mbox{\tiny $9$}}

\def\f14{\mbox{\tiny $\frac{1}{4}$}}

\def\R{\mbox{\tiny $R$}}

\def\B{\mbox{\tiny $B$}}

\def\mi{\mbox{\tiny $-$}}

\def\bb#1{\mbox{\footnotesize $(#1)$}}

\begin{document}

\title{Equilibrium and stability of neutrino lumps as TOV solutions}

\author{A. E. Bernardini}
\affiliation{Departamento de F\'{\i}sica, Universidade Federal de S\~ao Carlos, PO Box 676, 13565-905, S\~ao Carlos, SP, Brasil}
\email{alexeb@ufscar.br}

\date{\today}

\begin{abstract}
We report about stability conditions for static, spherically symmetric objects that share the essential features of mass varying neutrinos in cosmological scenarios.
Compact structures of particles with variable mass are held together preponderantly by an attractive force mediated by a background scalar field.
Their corresponding conditions for equilibrium and stability are given in terms of the ratio between the total mass-energy and the spherical lump radius, $M/R$.
We show that the mass varying mechanism leading to lump formation can modify the cosmological predictions for the cosmological neutrino mass limits.
Our study comprises Tolman-Oppenheimer-Volkoff solutions of relativistic objects with non-uniform energy densities.
The results leave open some questions concerning stable regular solutions that, to an external observer, very closely reproduce the preliminary conditions to form Schwarzschild black holes.
\end{abstract}

\date{\today}
\maketitle

\section{Introduction}

Theoretical issues concerning the understanding of physical processes where neutrinos are produced, scattered, or absorbed have been investigated through the years.
In parallel, strong arguments about the neutrino contribution to the energy density of the Universe have been constructed on very-well understood physical basis.
The role of neutrinos in the quintessence cosmology indeed comprises a major and still open issue that also involves the complete understanding of dark matter and dark energy, and the mechanism responsible for the onset of the accelerating phase of the Universe \cite{Zla98,Wan99,Ste99,Bar99,Ber00,Ame02,Kam02,Bil02,Ber02,Cal03,Mot04,Bro06A}.
In particular, models of mass varying neutrinos (MaVaN's) \cite{Far04,Pec05,Bja08,Ber08A,Ber08B,Han09,Pas09,Ber10} coupled to a dark energy light scalar field component have been considered for explaining  the emergence of an accelerating phase in recent cosmological times.

The simplest realization of the MaVaN mechanism consists in writing down an effective potential which, in addition to a scalar field dependent term, contains a term related to the neutrino energy density.
In the MaVaN scenarios the neutrinos remain essentially massless until recent times.
When their mass eventually grows close to its present value, they form a non-relativistic (NR) fluid and the interaction with the scalar field stops its evolution.
The potential energy of the dark energy component becomes the Universe's dominant contribution and the cosmological acceleration ensues.
However, MaVaN models face stability problems for the most of choices of neutrino-scalar field couplings and scalar field potentials.
Actually, when neutrinos become NR.
This instability is characterized by a negative squared speed of sound for an effective neutrino-dark energy coupled fluid, and results in the exponential growth of small scale modes \cite{Bea08}.
A natural interpretation for this is that the Universe becomes inhomogeneous with the neutrinos forming denser structures or lumps.
Effectively, the scalar field could mediate an attractive force between neutrinos leading to the formation of neutrino nuggets.

For the scenario of structure formation, we obtain static, spherically symmetric solutions of the Einstein equations that describe hypothetical neutrino lumps.
As a first approach, we study the Schwarzschild's so called interior solution for a sphere of incompressible fluid of constant density $\rho$ and pressure $p$ which drops from its central value to zero at the surface.
We determine the corresponding mass defect in order to obtain the stability conditions for neutrino lumps.
We also verify how the mass varying mechanism, as a mechanism to form compact structures, can modify the cosmological predictions for neutrino masses.
Due to the adiabatic conditions that we have set up, the thermodynamic pressure $p$ and its explicit dependence on $r$ are simply given in terms of $\rho\bb{r}$, and can be explicitly computed from the Tolman-Oppenheimer-Volkoff (TOV) equations for the hydrostatic equilibrium \cite{Tol39,Vol39}.
It justifies the assumption of an explicit analytical dependence of the variable mass on the radial coordinate of the massive body, $m\bb{r}$.
Actually, at our approach the neutrino mass behaves like an additional extensive thermodynamic degree of freedom.

The stability analysis reveals that our static solutions become dynamically unstable for modified Buchdahl limits \cite{Buc59} set by the ratio between total mass-energy and lump radius, $M/R$.
We also find regular solutions that to an outside observer resemble Schwarzschild black holes in formation.
It remains an open question whether such stable solutions do exist.

In section II, after briefly explaining the mass varying mechanism in the standard Friedman - Robertson - Walker (FRW) cosmological scenario, we establish the connection among the energy density $\rho$, the particle variable mass $m$ and the particle density $n$, for an adiabatic system of mass varying particles.
It results in a completely closed relation through which one can quantify the role of the mass varying mechanism in the study of structure formation \cite{Ber09F}.
We also summarize the procedure for obtaining the mass defect as a function of the lump radius and for performing the analysis of stability of equilibrium configurations.
In section III we obtain the conditions for neutrino lump formation in order to verify how the cosmological predictions for the neutrino masses could be changed.
Our analysis is extended to TOV solutions for matter lumps with non-uniform energy densities.
We draw our conclusions in section IV by summarizing our findings and discussing their implications.

\section{Equilibrium and stability in the mass varying scenario}

The influence of the spectrum of dynamical masses on thermodynamic of interactions are reported in the following.
In particular, we shall investigate the effects on stellar structures of neutral matter like, for instance, dark matter and {\em sterile}/active neutrinos \cite{Ber09E,Ber10}.
That is the reason for which we shall accomplish the idea of mutual interaction among neutral particles driven by the coupling to a scalar field through previously quoted mass varying mechanisms \cite{Far04,Bja08}.

\subsection{The mass varying mechanism}

In the usual mass varying scenario, mass varying particles are coupled to a light scalar field which is identified with the dark sector.
Presumably, the particle mass $m_{\nu}$ has its origin on the vacuum expectation value (VEV) of the scalar field so that its behaviour is governed by the dependence of the scalar field on the scale factor.
Given a particle statistical distribution $f\bb{q}$, in the flat FRW cosmological scenario, the corresponding energy density and pressure can be expressed by
\begin{eqnarray}
\rho\bb{a, \phi} &=&\frac{T^{\4}_{\0}}{\pi^{\2}\,a^{\4}}
\int_{_{0}}^{^{\infty}}{\hspace{-0.3cm}dq\,q^{\2}\, \left(q^{\2}+\frac{m^{\2}\bb{\phi}\,a^{\2}}{T^{\2}_{\0}}\right)^{\1/\2}\hspace{-0.1cm}f\bb{q}},\\
p\bb{a, \phi} &=&\frac{T^{\4}_{\0}}{3\pi^{\2}\,a^{\4}}\int_{_{0}}^{^{\infty}}{\hspace{-0.3cm}dq\,q^{\4}\, \left(q^{\2}+\frac{m^{\2}\bb{\phi}\,a^{\2}}{T^{\2}_{\0}}\right)^{\mi\1/\2}\hspace{-0.1cm} f\bb{q}},~~~~ \nonumber
\label{gcg01}
\end{eqnarray}
where $q \equiv \frac{|\mbox{\boldmath$p$}|}{T_{\0}}$, $T_{\0}$ is the background temperature at present, and we have assumed present-day values with $a_{\0} = 1$.

Simple mathematical manipulations allow one to demonstrate that
\begin{equation}
m\bb{\phi} \frac{\partial \rho\bb{a, \phi}}{\partial m\bb{\phi}} = (\rho\bb{a, \phi} - 3 p\bb{a, \phi}).
\label{gcg02}
\end{equation}
and, from the dependence of $\rho$ on $a$, one can obtain the energy-momentum conservation for the mass varying fluid,
\begin{equation}
\dot{\rho}\bb{a, \phi} + 3 H (\rho\bb{a, \phi} + p\bb{a, \phi}) =
\dot{\phi}\frac{d m\bb{\phi}}{d \phi} \frac{\partial \rho\bb{a, \phi}}{\partial m\bb{\phi}},
\label{gcg03}
\end{equation}
where $H = \dot{a}/{a}$ is the expansion rate of the Universe and the {\em overdot} denotes differentiation with respect to time ($^{\cdot}\, \equiv\, d/dt$).
The coupling between relic particles and the scalar field as described by Eq.~(\ref{gcg02}) is effective when particles are NR, i. e. $\frac{\partial \rho\bb{a, \phi}}{\partial m\bb{\phi}} \simeq n\bb{a} \propto{a^{\mi\3}}$ \cite{Far04,Bja08,Pec05}.
In opposition, as long as particles are relativistic ($T\bb{a} = T_{\0}/a >> m\bb{\phi\bb{a}}$), the decoupled fluid should evolve adiabatically since the strength of the coupling is suppressed by the relativistic pressure increasing ($\rho\sim 3 p$). The mass varying mechanism is essentially expressed by Eq.~(\ref{gcg02}), which translates the dependence of $m$ on $\phi$ into a dynamical behaviour.
In case of neutrinos, the most natural way to explain the smallness of the neutrino masses is through the seesaw mechanism, according to which, the tiny masses, $m$, of the usual left-handed neutrinos are obtained via a very massive, $M$, {\em sterile} right-handed neutrino for which the mass is driven by the VEV of a scalar field through a Yukawa coupling \cite{Ber08A,Ber08B}.

\subsection{Cosmon lumps}

The extremely weak interaction between dark matter and/or neutrinos and a slowly varying scalar field are sometimes described in terms of the {\em cosmon} dynamics \cite{Wet02,Wet08,Tet08,Ste96,Bil99}.
{\em Cosmon} lumps are bound objects, for which the scalar and gravitational fields combine to form non-linear solutions of their field equations.
The action of the {\em cosmon} coupled to mass varying particles ($\nu$ or DM) can be written as \cite{Wet08}
\begin{equation}
S = \int{\mbox{d}^{\4} x \sqrt{-g} \left[\frac{R}{16 \pi G} + \frac{1}{2}g^{\mu\nu}\partial_{\mu}\phi \partial_{\nu}\phi + V\bb{\phi} + \rho\bb{\phi}\right]},
\label{gcg03B}
\end{equation}
where $G$ is the Newton constant, $R$ is the Ricci curvature and $\rho\bb{\phi}$ is the corresponding mass varying particle energy density.
For static, spherically symmetric solutions, one employs the Schwarzschild metric given by $\mbox{d}s^{\2} = -B\bb{r} \mbox{d}t^{\2} + A\bb{r} \mbox{d}r^{\2} + r^{\2}\left(\mbox{d}\theta^{\2} + \sin^{\2}(\theta)\mbox{d}\varphi^{\2}\right)$, where $A = B^{-\1} = (1 - 2 G M /r)^{-\1}$ and $r$ is the radial coordinate.
From the field equations one can write
\begin{equation}
\frac{\mbox{d}^{2}\phi}{\mbox{d} r^{2}} + \left(\frac{2}{r} + \frac{B^{\prime}}{2 B} - \frac{A^{\prime}}{2 A}\right)
\frac{\mbox{d}\phi}{\mbox{d} r}
=
A \left (\frac{\mbox{d} V\bb{\phi}}{\mbox{d} \phi}  + \frac{\partial \rho\bb{\phi}}{\partial \phi} \right)
=
A \left (\frac{\mbox{d} V\bb{\phi}}{d \phi}  + \frac{\mbox{d} \ln{m\bb{\phi}}}{d \phi} (\rho - 3 p)\right)
\label{gcg03C}
\end{equation}
from which the explicit dependence of $\phi$ on $r$ is obtained.

Assuming the adiabatic approximation at cosmological scales \cite{Bea08}, the cosmological stationary condition \cite{Far04,Bro06A,Bja08,Ber08B} could be applied to Eq.~(\ref{gcg03C}) in order to suppress its right hand side.
Observing that the analytical dependence of the scalar field on $r$ should disappear far from the large concentration of matter, for $r$ values much larger than the surface radius $R$ one recovers the background flatness of a FRW Universe.
In this case, by adding the usual cosmological time dependent terms, the Eq.~(\ref{gcg03C}) should be rewritten as the usual energy conservation equation for the FRW Universe,
\begin{equation}
\ddot{\phi} + 3 H \dot{\phi} + \left (\frac{\mbox{d} V\bb{\phi}}{d \phi}  + \frac{\mbox{d} \ln{m\bb{\phi}}}{d \phi} (\rho - 3 p)\right)
= 0,
\label{gcg08}
\end{equation}
and the stationary condition should also be valid for a static approach \cite{Ber08A,Ber08B}, i. e.
\begin{equation}
\frac{\mbox{d} V\bb{\phi}}{d \phi}  + \frac{\mbox{d} \ln{m\bb{\phi}}}{d \phi} (\rho - 3 p)
= 0.
\label{gcg08C}
\end{equation}
The simplest scenario for this mechanism of structure formation is thus obtained from the extension of the cosmological stationary condition to the static configuration described by Eq.~(\ref{gcg03C}).
Introducing the abovementioned Schwarzschild expressions for $A\bb{r}$ and $B\bb{r}$, and assuming the stationary condition, Eq.~(\ref{gcg03C}) can be simplified as
\begin{eqnarray}
\frac{\mbox{d}^{2}\phi}{\mbox{d} r^{2}} + \left[\frac{2 - 8(M/R)(r^{\2}/R^{\2})}{r - 2(M/R) (r^{\3}/R^{\2})}\right]
\frac{\mbox{d}\phi}{\mbox{d} r}
&=& 0 ~~~~ (r < R),\nonumber\\
\frac{\mbox{d}^{2}\phi}{\mbox{d} r^{2}} + \left(\frac{2(r - M)}{r - 2 M}\right)
\frac{\mbox{d}\phi}{\mbox{d} r}
&=& 0 ~~~~ (r > R),
\label{gcg03CDM}
\end{eqnarray}
which gives
\begin{eqnarray}
\phi\bb{r} &=& \phi^{in}_{\0} + \frac{\phi^{in}_{\1}}{2M} \ln{\left(1 - \frac{2M}{r}\right)} ~~~~ (r < R),\nonumber\\
\phi\bb{r} &=& \phi^{out}_{\0} + \phi^{out}_{\1}\sqrt{(2M/R^{\3})}
\left[\frac{1}{\sqrt{(2 M /R^{\3})} r} - arc\tanh{(\sqrt{(2M/R^{\3})} r)}\right] ~~~~ (r > R),
\label{gcg03CDMA}
\end{eqnarray}
where $\phi^{in,out}_{\0,\1}$ are constants to be adjusted in order to match the boundary conditions.
In the Newtonian limit where $A\bb{r}\approx B\bb{r} \approx 1$, the constants match one each other so that $\phi^{in}_{\0,\1} \equiv \phi^{out}_{\0,\1} \equiv\phi_{\0,\1}$ and the above solutions reduce to
\begin{equation}
\phi\bb{r} = \phi_{\0} + \frac{\phi_{\1}}{r},
\label{gcg08D}
\end{equation}
which satisfies the suppression of the dependence on $r$, as one is away from the concentration of matter.

Since $\phi$ depends on the radial coordinate $r$ for generical classes of curved spaces some ordinary mass dependencies on $r$ can be tested.
Certainly the form of $m\bb{\phi}$ and the equation of state constraint may lead to quite different scenarios.
Thus, we emphasize that any prescription for dynamical masses is model dependent, i. e. arbitrary functions for $m\bb{r}$ are equivalent to arbitrary functions for $\phi\bb{r}$ and $m\bb{\phi}$.

\subsection{Thermodynamics of "non-interacting" mass varying particles}

To describe the connection among the extensive quantities $\rho$, $m$ and $n$ of an adiabatic system of mass varying particles in a stellar object, the thermodynamic equations can be summarized by
\begin{equation}
n \frac{\partial}{\partial n} = \rho  + p,
\label{star11}
\end{equation}
and
\begin{equation}
m \frac{\partial \rho}{\partial m} = \rho - 3 p.
\label{star12}
\end{equation}

The equation of state provides the dependence of pressure on energy density and specific entropy as $p = p\bb{\rho, s}$.
Due to the adiabatic conditions, the pressure and its explicit dependence on $r$ is simply given in terms of $\rho\bb{r}$ \cite{Ber09F}.
It can be explicitly obtained from TOV equations for the hydrostatic equilibrium.

Assuming that the equation of state for a symmetrically spherical distribution of matter where $p = p\bb{\rho\bb{r}}$ is determined by a univoquous dependence of $\rho$ on $r$, we have to find $\rho\bb{m\bb{r}, n\bb{r}, r}$ that satisfies the system of partial differential equations given by
\begin{equation}
\frac{\mbox{d}\rho}{\mbox{d}r} = \frac{\partial \rho}{\partial m}\frac{\mbox{d} m}{\mbox{d}r} + \frac{\partial \rho}{\partial n}\frac{\mbox{d} n}{\mbox{d}r} + \frac{\partial\rho}{\partial r},
\label{star13}
\end{equation}
and Eqs.~(\ref{star11})-(\ref{star12}).
Provisorily eliminating $p$ from Eqs.~(\ref{star11})-(\ref{star12}), we obtain
\begin{equation}
4 \rho = m \frac{\partial}{\partial m} + 3 n \frac{\partial}{\partial n},
\label{star14}
\end{equation}
and consequently, once setting the variable dependence $\alpha \rightarrow \alpha\bb{r}$,
\begin{equation}
\rho\bb{m, n} = \kappa m^{\1-\3 \alpha} n^{\1 + \alpha}
\label{star15}
\end{equation}
satisfies Eq.~(\ref{star13}).
Identifying $\alpha\bb{r}$ as $p\bb{r}/\rho\bb{r}$, one finds
\begin{equation}
\rho = \rho\bb{m\bb{r}, n\bb{r}, r} = m^{\1-\3 \frac{p}{\rho}} n^{\1 + \frac{p}{\rho}},
\label{star16}
\end{equation}
where, for simplicity, the explicit dependence of $m$, $n$, $p$ and $\rho$ on $r$ was omitted, the constant $\kappa$ was adjusted to have $\rho = m \, n$ in the NR limit, and $\rho = n^{\4/\3}$ in the ultra-relativistic (UR) limit.
Obviously, the explicit form of $\rho\bb{r}$ takes the place of an equation of state, since $p\bb{r}$ can be determined from hydrostatic equilibrium, i. e. from TOV equations \cite{Tol39}.

The above results were applied for obtaining the stability conditions of symmetrically spherical lumps of mass varying particles assuming them as incompressible structures of constant energy density $\rho_{\0}$ \cite{Ber09F}.

\subsection{Equilibrium and stability}

In the relativistic domain, we have
\begin{equation}
M = M_{\0} + M_{\B} = 4 \pi \int_{\0}^{\R}{r^{\2} \rho\bb{r}\,\mbox{d}r},
\label{star18A}
\end{equation}
where
\begin{equation}
M_{\0} = 4 \pi \int_{\0}^{\R}{r^{\2} \, m\bb{r} n\bb{r}\, \sqrt{A\bb{r}}\,\mbox{d}r},
\label{star18B}
\end{equation}
with $c = 1$.

The form of Eq.~(\ref{star18A}) suggests that one interprets $M$ as the total (mass) energy inside a radius $R$, including the rest mass-energy $M_{\0}$, the internal energy of motion $W$ and the (negative) potential energy of self-gravitation $U$.
By following the {\em auxiliary} equation given by Eq.~(\ref{star12}), $M$ is reduced to $M_{\0}$ (no binding energy and no motion) if $p = 0$.
The difference between the total rest mass-energy and the total energy corresponds to the (positive) binding energy $M_{\B}$ which keeps the stellar structure stable,
\begin{equation}
M_{\B} =  M_{\0} - M = 4 \pi \int_{\0}^{\R}r^{\2}{\left[ \rho\bb{r} - m\bb{r} n\bb{r}\sqrt{A\bb{r}}\right]\mbox{d}r}.
\label{star18C}
\end{equation}

Although it is very useful to interpret $M$ as the total energy, the internal energy of motion $W$ and the (negative) potential energy of self-gravitation $U$ are not particularly useful except in the Newtonian approximation where the approximation $M_{\B} \approx U + W$ is realistic.
Otherwise, the rest mass-energy $M_{\0}$ is fundamental for the stability analysis.
The binding energy is sometimes called {\em mass defect} and it corresponds to the energy released during the formation of a dense star from initially rarefied matter: a typical mechanism for lump formation.
Because of this we have $M_{\B} > 0$ for a completely stable static body originating from diffuse matter.

In \cite{Ber09F} we have shown how the mass varying mechanism could modify the equilibrium condition based on the mass defect criterium for relativistic stellar objects.
Two interesting effects can be achieved from such a previous analysis: one for particles with increasing mass inwards to the center, and the other for particles with decreasing mass.
Observing the mass defect for compact objects composed by particles with their mass exponentially increasing inwards to the center, we notice that stable configurations are tremendously favored.
Decreasing mass inwards to the center can also form stable structures up to certain limiting values for $M/R$.
The equilibrium conditions are achieved just for small values of $M/R$, conditioning the existence of neutrino lumps to compact objects with arbitrarily small masses.
The crucial point in \cite{Ber09F} is that coupling with the background scalar field is relevant in determining the limits for stability.

In the absence of mass varying mechanisms the rest mass-energy $M_{\0}$ should be  written as $M_{\0} = m \, N$, where $N$ is the total number of particles inside the radius $R$,
\begin{equation}
N = 4 \pi \int_{\0}^{\R}{r^{\2} n\bb{r} \sqrt{A\bb{r}}\,\mbox{d}r}.
\label{star19}
\end{equation}



It is important to notice that, for the static approach, the stability analysis is the same as when we treat mass varying particle systems with the thermodynamics governed by Eq.~(\ref{star16}).
As previously quoted \cite{Ber09F}, the only difference is that the analysis has to be performed in terms of $M_{\0}$ (in confront with $M$) and not in terms of $N$.
It is translated by the correspondence between $M$ and $M_{\0}$ in terms of $\mbox{d}M/\mbox{d}M_{\0}$.
For stable equilibrium configurations \cite{ZelXX,ThoXX} one always has $\mbox{d}M/\mbox{d}M_{\0} < 1$.
The stability condition is reduced to $\mbox{d}M/\mbox{d}N < m$ for the case where the particle masses does not depend on the radial coordinate, $m\bb{r} = m = m_{\0}$.
Because of this, the usual general relativity analysis of stability is sometimes performed in terms of $M$ {\em versus} $N$.
In case of mass varying particle systems it causes substantial deviations from the accurate scenario described in terms of $M$ {\em versus} $M_{\0}$.

The relation between the total (mass) energy $M$ and the rest mass-energy $M_{\0}$ and its correspondence with the precedent analysis in terms of the total number of particles $N$ bounded by the surface of radius $R$ was obtained in \cite{Ber09F}.
The stable equilibrium configurations are achieved when $\mbox{d}M/\mbox{d}M_{\0} < 1$, and not when $\mbox{d}M/\mbox{d}N < 1$.

\section{Neutrino lumps with non-uniform energy density}

In the usually proposed MaVaN scenarios \cite{Far04,Ber08A,Ber08B,Ber09E,Ber10} the neutrinos remain essentially massless until recent times.
When their mass eventually increases close to its present value and the interaction with the background scalar field almost ceases \cite{Ber08A,Ber08B}.
The energy of the scalar field becomes the dominant contribution to the energy density of the Universe.
Cosmological acceleration ensues.
For the coupled neutrino-scalar field fluid the squared speed of sound may become negative - a signal of instability \cite{Bea08}.
The natural interpretation of this is that the Universe becomes inhomogeneous with neutrino overdensities subject to nonlinear fluctuations \cite{Mot08} which eventually collapses into compact neutrino lumps.

Unlike photons and baryons, cosmological neutrinos have not been observed, so arguments about their contribution to the total energy density of the Universe are necessarily theoretical.
The question posed here is if the fraction corresponding to the neutrino energy contribution is due to an isotropic and homogenous NR ($p \sim 0$) distribution of particles, $M_{\0}$, or if it is due to a gas of weakly interacting particles ($p > 0$) forming lumps in an expanding Universe, $M$.
The ratio given by $M_{\0}/M$ for different scenarios of mass varying mechanism lead to a reasonable estimative for corrections on the absolute values of the neutrino masses.

Current cosmological data constrain the number of active neutrino flavours as well as the sum of their masses to $\displaystyle\sum_{i = \1}^{\3} m_{\nu i}\,  <\,  0.75 \,eV$ at $95\,\%$  c.l. \cite{Han05}.
This constraint does not agree with the Heidelberg-Moscow bounds arising from of neutrinoless double beta decay which sets the limit $\displaystyle\sum_{i = \1}^{\3} m_{\nu i}\,  >\,  1.2 \,eV$ at $95\,\%$  c.l. \cite{Kla04}, which, however, are in agreement with the CMB data analysis of the WMAP results that sets $0.7 \,eV$  at $95\,\%$ c. l. for each neutrino species (and $2.0\, eV$ in total) \cite{Ich06}.
Actually, improvement on experimental data are expected to be sensitive to the effects of a finite sum of neutrino masses as small as $0.06\, eV$ \cite{Han05,Pas07}, the lower limit arising by neutrino oscillation experiments that set $\Delta m^{\2} \sim  7.5$ - $8.7 \times 10^{\mi\5} \, eV^{\2}$ ($2 \sigma$) for solar neutrinos, and $\Delta m^{\2} \sim  1.7$ - $2.9 \times 10^{\mi\3} \, eV^{\2}$ ($2 \sigma$) for atmospheric neutrinos.

Meanwhile, all the cosmological neutrino mass predictions are performed for an isotropic and homogenous NR distribution of particles, which in the total absence of internal interactions and gravitational forces result in a total energy (per equivalent lump) equal to $M_{\0}$.
For the case where the total neutrino number $N$ is assumed to be conserved, one can define the apparent (measured) value for neutrino masses as $\langle \mu \rangle = M/N$ and the realistic (expected) corresponding value as $\langle m \rangle = M_{\0}/N$ so that
\begin{equation}
\frac{\langle m \rangle}{\langle \mu \rangle} = \frac{M_{\0}}{M}.
\label{star20}
\end{equation}
Let us then assume that the energy density of the Universe involves a gas of weakly interacting particles (neutrinos) conceived by the previously considered static, spherically symmetric solution of the Einstein equation with a constant energy density.
Independently of the analytical characteristic of the mass varying mechanism, we assume that the neutrino mass is equal to $m_{\0}$ on the surface of the lump, where $r = R$ and $p = 0$.

\begin{figure}
\vspace{-1.0 cm}
\centerline{\psfig{file=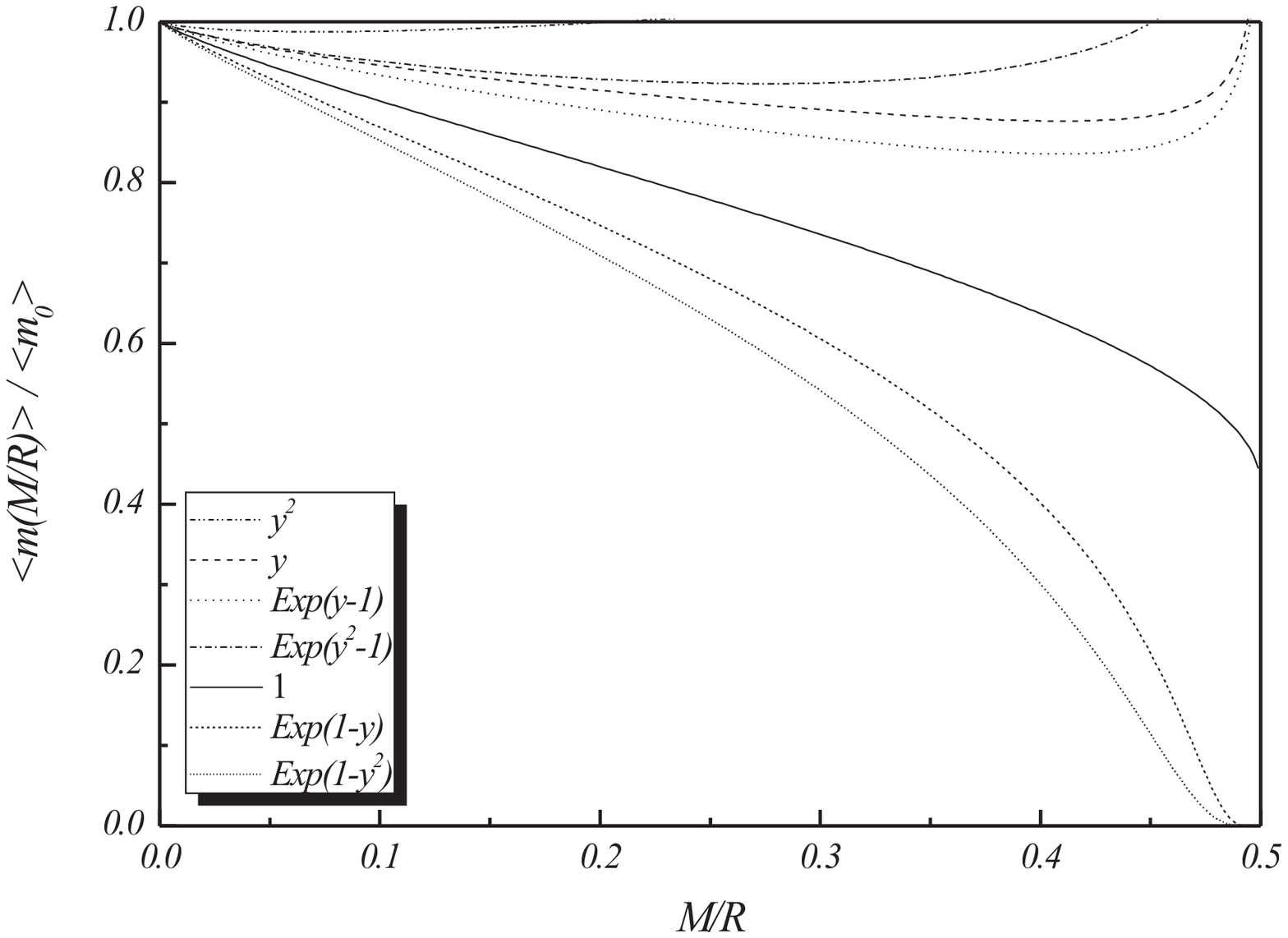, width=14cm}}
\vspace{-1.0 cm}
\caption{Modified predictions for the absolute value of neutrino masses due to the formation of neutrino lumps for several {\it ad hoc} mass varying particle scenarios for compact objects with uniform energy density \cite{Ber09F}.
The preliminary description for the dynamical mass dependence on $y = r/R$, is given on the rectangle.
For all stable scenarios the absolute neutrino mass predictions are shifted towards smaller values.}
\label{Fstar07}
\end{figure}

In the Fig.~\ref{Fstar07} we represent the ratio $\langle m \rangle/\langle \mu \rangle$ for different mass varying scenarios in order to obtain the realistic value of neutrino masses in comparison with the predicted values.
It corresponds to the stability curves obtained in \cite{Ber09F}.

Obviously, the smaller is the spherical radius, the more isotropic and homogeneous is the cosmological neutrino distribution.
For lumps in which the neutrino mass decreases from the surface to the center, we have a lower limit for $\langle m \rangle/\langle \mu \rangle$.
It coincides with the upper limit for the radius of stable equilibrium configurations.
For the other cases where the neutrino mass increases inwards to the center, we have completely stable equilibrium configurations for which $\langle m \rangle/\langle \mu \rangle$ decreases up to vanish on the surface of radius $R$.
From the analysis of equilibrium and stability, we have noticed that an unlimited quantity of (mass) energy $M$ can be compacted as a lump when the particle mass increases inwards to the center of the star, constituting the preliminary conditions to form a kind of black-hole where the limitrophe radius correspond to the event horizon.
A crude interpretation of this is that neutrinos forming black-holes can be practically non-massive at the their surfaces.
Although it seems paradoxical, it can be read as a natural consequence of the mass varying mechanism.
Actually, for these compact lumps, the most part of its mass is concentrated in the center.
Assuming that its energy contribution to the cosmological scenario and the total number of particles are the same (there being or there not being lumps!), the absolute mass value vanishes on the surface of these objects if the Buchdahl's limit is reached.

In general, stable scenarios lead to a shift towards smaller values for the absolute neutrino mass predictions.
Comparing two possible scenarios for cosmological neutrinos coupling with a background scalar field, one with an isotropic and homogenous energy distribution, and the other with perturbations which results in stable neutrino lumps, the presence of neutrino lumps in the latter case should bring the absolute neutrino mass predictions to
lower values, even with $m_{\0}\lesssim 0.07 \, eV$, which should be problematic.

Obtaining explicit analytical solutions of Einstein's gravitational field equations is not a trivial issue, on account of their complicated and nonlinear character.
Even for the simple case of static gravitational equilibrium for a spherically symmetric matter distribution with uniform energy density, there are only two explicit solutions of the TOV equations: the solution for the Einstein's Universe and the Schwarzschild solution studied in \cite{Ber09F}.

Our purpose is to verify how relativistic structures with non-uniform energy density described by TOV solutions have their equilibrium and stability conditions modified by the mass varying mechanism.
We are focused on solutions to use in investigating fluids with infinite density and pressure at the center of the spherical distribution of matter.
We consider the following two well-known TOV solutions \cite{Tol39},

\paragraph{Solution A}
\begin{eqnarray}
A\bb{r/R,\,\kappa} &=& \left[\frac{1+2\kappa-\kappa^{\2}}{1-(r/R)(1+2\kappa-\kappa^{\2})\left(\frac{\kappa^{\2}}{(1+2\kappa-\kappa^{\2})(1+2\kappa)}\right)^{\frac{1+2\kappa-\kappa^{\2}}{2\kappa-\kappa^{\2}}}}\right]^{\frac{\1}{\2}},\nonumber\\
8 \pi \rho\bb{r/R,\,\kappa} &=& \frac{2\kappa-\kappa^{\2}}{1+2\kappa-\kappa^{\2}}\frac{1}{(r/R)^{\2}} + \frac{3+5\kappa-2\kappa^{\2}}{1+\kappa}\frac{\kappa^{\2}}{(1+2\kappa-\kappa^{\2})(1+2\kappa)}(r/R)^{\frac{2\kappa-2\kappa^{\2}}{1+\kappa}},\nonumber\\
8 \pi p \bb{r/R,\,\kappa} &=& \frac{\kappa^{\2}}{1+2\kappa-\kappa^{\2}}\left(\frac{1}{(r/R)^{\2}} - (r/R)^{\frac{2\kappa-2\kappa^{\2}}{1+\kappa}}\right).
\label{star21}
\end{eqnarray}
for which $M/R = \kappa/(1-2\kappa)$.

In this case, the ratio of the total mass $M$ to the radius $R$ cannot reach values greater than $1/4$ if one assumes that $\rho\bb{0} > 3 p\bb{0}$, i. e. the {\em soft} Buchdahl's limit.

\paragraph{Solution B}
\begin{eqnarray}
A\bb{r/R,\,\kappa} &=& \left(2-\kappa^{\2}\right)^{\frac{\1}{\2}},\nonumber\\
8 \pi \rho\bb{r/R,\,\kappa} &=& \frac{1-\kappa^{\2}}{2-\kappa^{\2}}\frac{1}{(r/R)^{\2}},\nonumber\\
8 \pi p \bb{r/R,\,\kappa} &=& \frac{\left(1-\kappa^{\2}\right)^{\2}}{2-\kappa^{\2}}\frac{1}{(r/R)^{\2}}
\frac{1-(r/R)^{\2 \kappa}}{(1+\kappa)^{\2} - (1-\kappa)^{\2}(r/R)^{\2 \kappa}}.
\label{star22}
\end{eqnarray}
for which $M/R = (1 - \kappa^{\2}/(4-2\kappa^{2})$.

In this case, the ratio of the total mass $M$ to the radius $R$ cannot reach values greater than $3/14$ if one assumes that $\rho\bb{0} > 3 p\bb{0}$.

\begin{figure}
\vspace{-1.0 cm}
\centerline{\psfig{file=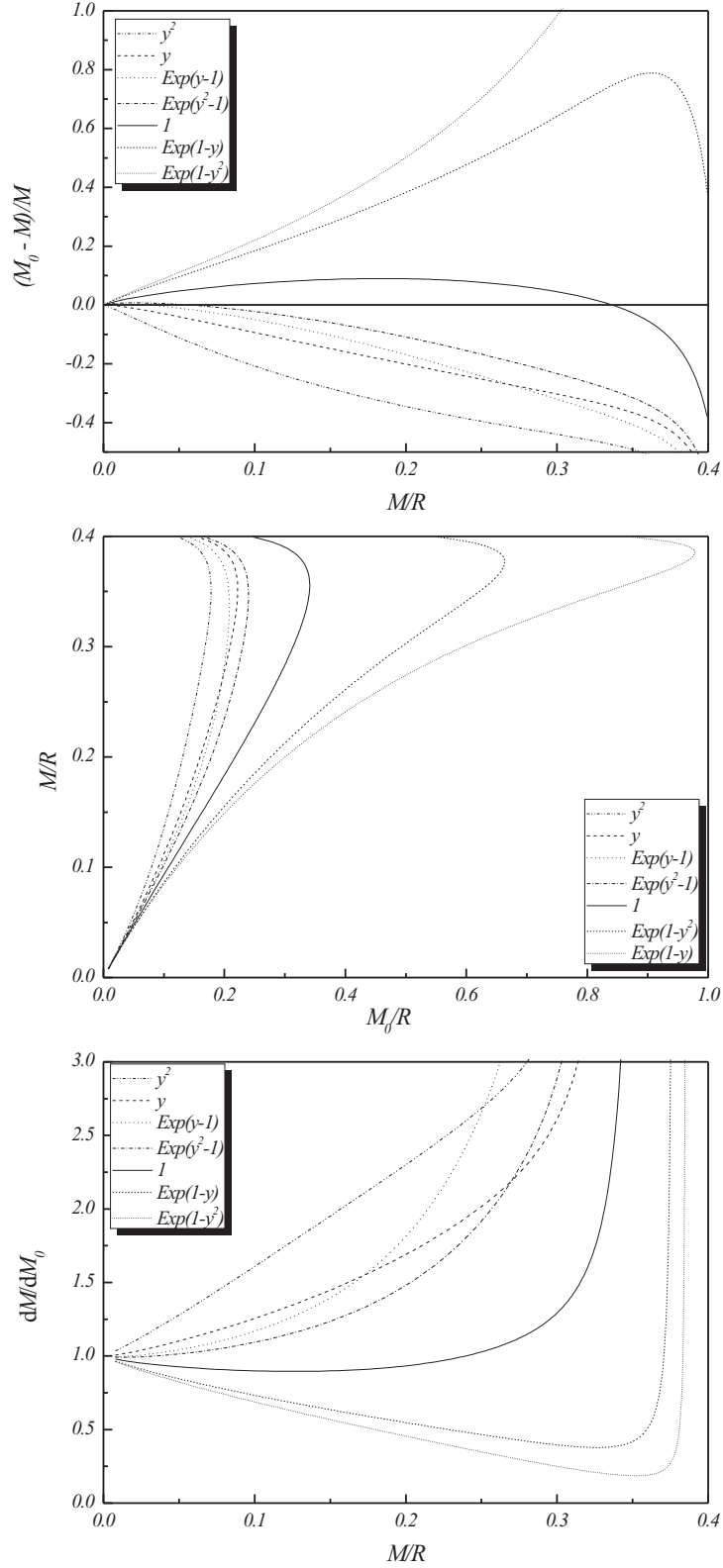, width=12cm}}
\vspace{-1.0 cm}
\caption{The mass defect, the stability curves and the stability conditions for compact objects (relativistic stars) with variable density described by TOV solution A.
The correspondence with the mass varying mechanism prescription is maintained.
At this point, we restrict our analysis to the accurate procedure which takes into account $M$ and $M_{\0}$, and discards $N$.}
\label{Fstar08}
\end{figure}

\begin{figure}
\vspace{-1.0 cm}
\centerline{\psfig{file=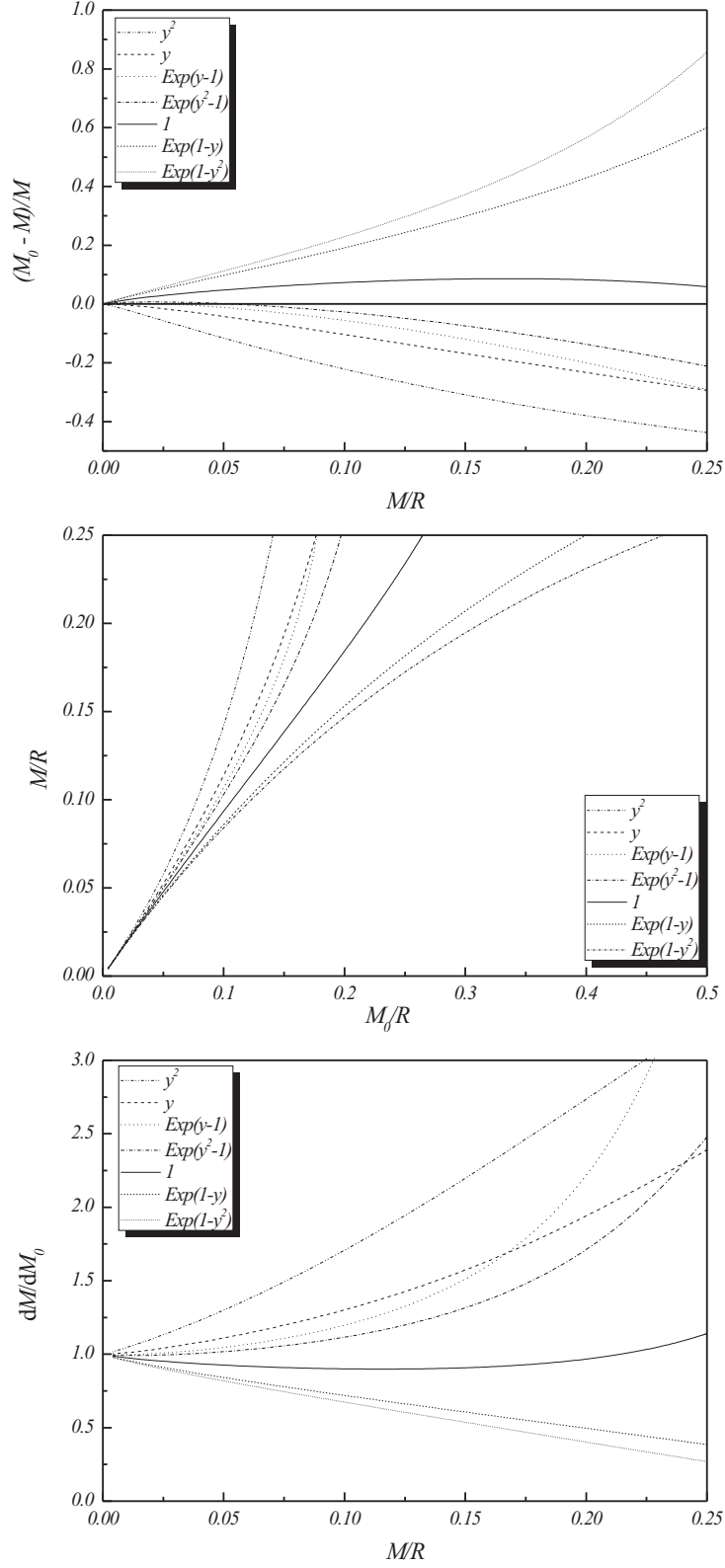, width=12cm}}
\vspace{-1.0 cm}
\caption{The mass defect, the stability curves and the stability conditions for compact objects (relativistic stars) with variable density described by TOV solution B.}
\label{Fstar09}
\end{figure}

By following an analogous procedure as that of \cite{Ber09F}, in the Figs.~\ref{Fstar08} and \ref{Fstar09} we quantify the equilibrium and stability conditions for the above TOV solutions.

A complete similarity between the two sets of solutions can be noticed.
Despite being difficult to observe, the gravitationally unstable configurations for Solution A, in comparison with Solution B, are smoothly shifted toward larger values of the ratio $M/R$.
In comparison with the objects with uniform energy density, the binding energy and the stability limits, in terms of the ratio $M/R$ are severally attenuated.
The crucial difference between constant and variable energy densities concerns the form of the solutions with large values of the ratio $M/R$.
The attractive interaction mediated by the scalar field in both cases, A and B, are not sufficiently strong to compact structures with large masses.
Gravity must play a more relevant role for these compact objects to exist.

For the four cases of mass dependence on $r$ where we have neutrino masses decreasing inwards to the center, a striking feature is the existence of neutrino lumps with arbitrarily small masses.
They correspond to the lower left corner of the figures, where both, the mass defect and the ratio $M/R$ vanish.
Because of the small masses of this objects, the contribution from gravity is negligible and their existence is a consequence of the attractive force mediated by the scalar field.
The reason is that the neutrinos are essentially massless inside the compact structures.
For the other three cases, which includes lumps of neutrinos with constant mass, the stability conditions can also be achieved for massive objects with larger values of $M/R$.
However, the increasing of the binding energy is limited by the {\em soft} Buchdahl's limit.
It corresponds to an exotic equation of state at the center of the object, i. e. $\rho\bb{0} > 3 p\bb{0}$.
Differently from which we have observed in the analysis of structures with uniform energy density,
the conditions to form structures analogous to black-holes are not provided here.

\begin{figure}
\vspace{-1.0 cm}
\centerline{\psfig{file=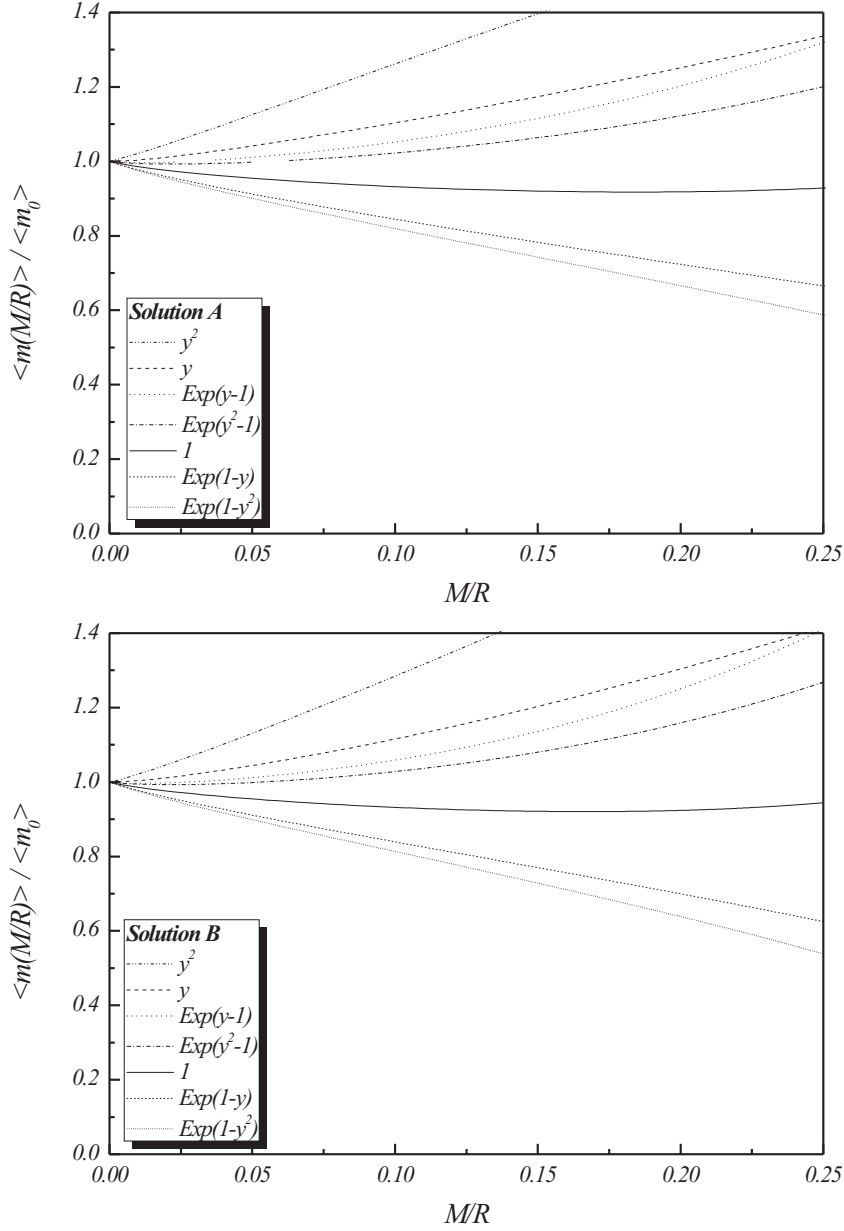, width=14cm}}
\vspace{-1.0 cm}
\caption{Modified predictions for the absolute value of neutrino masses due to the formation of neutrino lumps.
We plot the ratio $\langle m \rangle/\langle \mu \rangle$, where $\langle m \rangle$ corresponds to the absolute mass value for neutrinos on the surface of the compact object, i. e. $r = R$, and $p = 0$.
We compare the predictions for type A and type B TOV solutions for relativistic stars with variable energy density.
We plot them in the same scale ($M/R \in (0, 0.25)$) in order to verify that the modifications from one to another are minimal.}
\label{Fstar1011}
\end{figure}

With respect to the consequences of the neutrino lump formation in how the population of lumps can modify the cosmological predictions for the neutrino masses, we have the scheme of the Fig.~\ref{Fstar1011}.
For non-uniform densities here considered, the modifications to the mass predictions accomplish the stability conditions for structure formation.

\section{Conclusions}

We have analyzed the stability conditions for static, spherically symmetric solutions of the Einstein equations for a system of non-baryonic matter which forms stable structures due to attractive forces mediated by a background scalar-field (dark energy).
We have supposed that these astrophysical objects share the key features of the mass varying mechanism in cosmological scenarios.
In case of dark matter scenarios, the typical size of these overdensity fluctuations is supposed to be large, in the range of superclusters and beyond.
The natural interpretation of these instabilities is that the Universe becomes inhomogeneous with neutrino overdensities subject to nonlinear fluctuations that eventually collapses into neutrino lumps.
In effective terms, a scalar field  mediates the attractive force among the constituting particles leading to the formation of stable lumps.
Actually, this would turn the combined fluid to form compact structures which behave like cold dark matter.

Our analysis leaves open the question whether compact structures formed by neutrinos, as regular and stable solutions of TOV equations, indeed exist.
After reporting about the equilibrium and stability conditions for relativistic compact structures with uniform energy density, we have extended the calculations to other two sets of TOV solutions.
We have established the connection between the population and size of neutrino lumps and the modification on the standard cosmological predictions for absolute values of the neutrino mass.

In addition, two distinct effects could be achieved from the study of the mass defect for compact objects.
One for particles with increasing mass inwards to the center, and the other for particles with decreasing mass.
For compact objects composed by particles with mass exponentially increasing inwards to the center, stable configurations are naturally obtained.
In particular, compact structures with a uniform energy density has the binding energy increasing to infinite as the ratio $M/R$ approximates to the Buchdahl's limit.
When these objects shrink to such size they must inevitably keep shrinking, and eventually form a black hole.
That is not the case of compact objects with non-uniform energy densities, where the conditions to form structures analogous to black-holes are not completely provided.
In the same way,  up to certain boundary values for $M/R$, decreasing mass inwards to the center can lead to stable structures.
Differently from the former results, the equilibrium conditions are achieved just for smaller values of $M/R$, constraining the existence of neutrino lumps to compact objects with arbitrarily small masses.
In fact, the coupling with the background scalar field is relevant in determining the bounds for stability.
At the same time, the possibility of neutrino lumps to follow their collapse essentially due to the scalar mediated attractive interaction has to be more carefully elaborated.

We have assumed that the radius and mass of the compact objects which belong to the family of allowed TOV solutions depend on details of the dynamical formation mechanism.
As we have quoted before, the subtleties of the study of the neutrino mass generation mechanism and the analytical dependence of the scalar field on the spacetime curvature are not mandatory.
Since the Higgs sector \cite{Ber09,Ber09C,Ber09E,Ber10} and the neutrino sector are possibly the only ones where one can couple a new standard model (SM) singlet without upsetting the known phenomenology, the replacement of the explicit dependence of the neutrino mass on the scalar field $\phi$ by a direct link to the radial coordinate of the curved space is acceptable.

To end up, we emphasize that allowing for a dynamical behaviour to a scalar field associated to dark energy in connection with the SM neutrinos and the electroweak interactions may bring important insights on the physics beyond the SM.
Neutrino cosmology, in particular, is a fascinating example where salient questions concerning SM particle phenomenology can be addressed and hopefully better understood.
\begin{acknowledgments}
The author thanks for the financial support from the Brazilian Agencies FAPESP (grant 08/50671-0) and CNPq (grant 300627/2007-6).
\end{acknowledgments}

\pagebreak
\newpage

\end{document}